\documentclass[10pt,preprint]{article}
\usepackage{amssymb}
\usepackage{amsmath}
\usepackage{graphics}
\usepackage{graphicx}
\usepackage{epsfig}

\renewcommand{\vec}[1]{{\bf #1}}
\newcommand{\eqb}{\begin{equation}}
\newcommand{\eqe}{\end{equation}}
\newcommand{\dmb}{\begin{displaymath}}
\newcommand{\dme}{\end{displaymath}}

\newcommand{\eab}{\begin{eqnarray}}
\newcommand{\eae}{\end{eqnarray}}

\newcommand{\e}{\mbox{e}}
\newcommand{\be}{\begin{equation}}
\newcommand{\ee}{\end{equation}}

\begin{document}

\begin{titlepage}
\begin{flushright} 
.
\end{flushright}
\vspace{0.2cm}
\begin{center}
\Large{Spatial Wilson loop in continuum, deconfining SU(2) Yang-Mills thermodynamics}
\vspace{0.5cm}\\ 
\large{Josef Ludescher$^*$, Jochen Keller$^*$, 
Francesco Giacosa$^\dagger$, and\\ 
Ralf Hofmann$^{*}$}
\end{center}
\vspace{0.5cm} 
\begin{center}
{\em $\mbox{}^*$ Institut f\"ur Theoretische Physik\\ 
Universit\"at Heidelberg\\ 
Philosophenweg 16\\ 
69120 Heidelberg, Germany}
\end{center}
\vspace{0.1cm}
\begin{center}
{\em $\mbox{}^\dagger$ Institut f\"ur Theoretische Physik\\ 
Universit\"at Frankfurt\\ 
Johann Wolfgang Goethe - Universit\"at\\ 
Max von Laue--Str. 1\\ 
60438 Frankfurt, Germany}
\end{center}
\vspace{2.8cm}
\begin{abstract}

The uniquess of the effective actions describing 4D SU(2) and
SU(3) continuum, infinite-volume Yang-Mills thermodynamics in their deconfining 
and preconfining phases is made explicit. Subsequently, 
the spatial string tension 
is computed in the approach proposed by
Korthals-Altes. This SU(2) calculation is 
based on a particular, effective two-loop correction to 
the pressure needed for the extraction of the hypothetic number density
of isolated and screened magnetic monopoles or antimonopoles in the
deconfining phase. 
By exponentiating the exchange of the tree-level massless but one-loop
dressed mode within a quadratic
spatial contour of side-length $L$ in the effective theory we demonstrate that for $L\to\infty$
the Wilson loop exhibits {\sl perimeter} law. This is in contrast to a rigorous 
{\sl lattice} result subject to the Wilson action and for this action 
valid at sufficiently high temperature. In the framework of the effective theory 
there is, however, a regime for small
(spatially unresolved) $L$ were the exponent of the spatial Wilson loop 
possesses curvature as a function of $L$. 

\end{abstract} 

\end{titlepage}

\section{Introduction\label{intro}}

In the past the pursuit of a nonperturbative 
understanding of 4D Yang-Mills theory at high 
temperature was limited to lattice formulations. A prominent
target of lattice investigations is the spatial Wilson loop evaluated on a
rectangular and planar contour $C$. Based on an analogy to the 
strong-coupling expansion, whose validity is suggested at sufficiently
high temperature by the part of the Wilson action containing 
purely spacelike plaquettes, it was shown in \cite{Borgs1985} 
that at a given, finite, spatial lattice spacing the spatial Wilson loop always
exhibits area law implying the existence of a spatial string tension
$\sigma_s$. For the gauge group SU(2), which we will mainly be concerned
with here, this 
result was verified subsequently by direct 
simulation \cite{Polonyi1987,Bali1993,spatialWL}. 

As already pointed out in \cite{Polonyi1987} the argument made 
in \cite{Borgs1985} neither has much to say about the cutoff dependence of 
$\sigma_s$ in the ultraviolet nor about the behavior of $\sigma_s$ 
in the infinite-volume limit since it
inherently relies on a finite spatial discretization allowing, at 
leading order in the strong-`coupling'-expansion, to 
tile the minimal area spanned by $C$ in terms of spacelike 
plaquettes. This then implies the area law in the lattice 
formulation.

In contrast to lattice formulations, 
the nonperturbative approach to SU(2) and SU(3) Yang-Mills 
thermodynamics developed in 
\cite{Hofmann2005,HofmannGiacosa2006,HofmannLE2006,Hofmann2007} uniquely derives an
effective theory for the deconfining phase starting from a genuine
spacetime continuum, space being infinitely extended, and from the
fundamental (euclidean) Yang-Mills action $S_{\tiny\mbox{YM}}$. The effective theory 
emerges as a consequence of pursuing the following chain of steps: 
1) Consider the BPS saturated (topologically nontrivial) sector of (periodic) euclidean field 
configurations (time variable imaginary). Independent of their topological charge 
these (anti)selfdual configurations possess vanishing energy-stress. 
This implies that 2) whatever effective field is obtained as a consequence 
of performing a spatial coarse graining over them it is {\sl nonpropagating}. 
But a nonpropagating effective field acts as a static background. 3) Such a background configuration would break the spatial isotropy and/or homogeneity of the thermal system (one-point functions) unless it is a scalar 
field. That is, no Lorentz tensor of rank greater than zero may emerge as an effective field from the sector 
of fundamental BPS saturated field configurations. If this effective scalar would be neutral 
under the gauge group then it would not participate in the thermodynamics due 
to its decoupling from the $Q=0$ sector of propagating field configurations and its vanishing effective energy-stress. 4) Since this scalar field does 
not propagate its classical equation of motion is, in a given gauge, 
solely determined by the euclidean time dependence of its phase. 5) It turns out \cite{HerbstHofmann2004,Hofmann2007} that only a single nonlocal 
definition of this time dependence, involving an adjointly transforming two-point function of the fundamental field strength, is possible. 6) This definition only admits the contribution of configurations with topological charge modulus $|Q|=1$ (Harrington-Shepard (anti)calorons, trivial holonomy) because of too many dimensionful parameters spanning the moduli spaces\footnote{Due to the temperature independence 
of the weight $\e^{8\pi^2|Q|/g^2}$ ($g$ the fundamental, classical and thus temperature independent coupling {\sl constant} the definition of the time dependence of the 
field's phase must not invoke any explicit temperature dependence. This only leaves 
room for the spatial coarse-graining over a two-point function of the fundamental field strength.} of higher-charge calorons and the 
instability of (anti)calorons of nontrivial holonomy \cite{Diakonov,GPY1981}. 
7) Consistency of the thus derived second-order equation and the effective BPS saturation (first-order equation) of the adjoint scalar field uniquely fixes its gauge invariant 
potential with the Yang-Mills scale $\Lambda$ entering as a (nonperturbative) integration 
constant \cite{HofmannGiacosa2006, Hofmann2007}. 8) Since no field other than the inert 
adjoint scalar may emerge from the sector(s) of BPS saturated fundamental 
field configurations the coupling to the sector of propagating gauge fields 
($Q=0$), which by perturbative renormalizability \cite{tHooftVeltman} 
by itself cannot generate higher dimensional operators upon the invoked 
spatial coarse-graining, occurs minimally via the square of the adjoint 
covariant derivative, see also Sec.\,\ref{Uni}. 9) The thus derived effective 
action yields an accurate a priori estimate for the thermal ground state 
by solving the Yang-Mills equation of the effective $Q=0$ sector 
in the background of the adjoint scalar field. 10) Radiative corrections to the free quasiparticle situation described by the effective action can be integrated out in a rapidly converging loop expansion subject to constraints imposed by the existence of an effective thermal ground state \cite{HofmannLE2006}. Interestingly, only planar bubble 
diagrams contribute. As we will see, the maximal resolution of the 
effective theory, which is given by the scalar field's modulus, almost 
everywhere in temperature is too small to allow for the existence and dynamics of single magnetic monopoles and antimonopoles, generated by (anti)caloron deformation away from trivial holonomy and away from BPS saturation, to be resolved.  

We consider it useful to repeat the above steps including physical interpretations. 
A sufficiently local spatial coarse-graining \cite{HerbstHofmann2004} 
over stable and absolute minima of the 
action in the sector with topological charge modulus $|Q|=1$
(Harrington-Shepard (anti)calorons \cite{HS1977}, nonpropagating
configurations) generates a spatially homogeneous and inert adjoint
scalar field $\phi$ whose modulus sets the maximally available 
resolution in the effective theory\footnote{A nontrivial-holonomy
  (anti)caloron \cite{NahmVanBaalLeeLu} is not
  stable. Moreover, it is suggested to consider a slow dependence in real time of 
the holonomy parameter $u$ since a constancy of $u$ in space and time would completely
suppress the weight of the nontrivial-holonomy caloron in the thermodynamic limit
\cite{GPY1981}. This time dependence, however, necessarily lifts the associated field 
configuration above the BPS bound. On the level of the effective theory,
where the thermal ground state is described by the field $\phi$ together
with a pure-gauge configuration of the coarse-grained $Q=0$ sector, this
microscopic lift above the BPS bound is seen in terms a positive ground-state
energy density which is the negative of its pressure
\cite{Hofmann2005,Hofmann2007}.}. Recall, that only this
sector is allowed to contribute to the kernel containing $\phi$'s
phase \cite{Hofmann2005,Hofmann2007,HerbstHofmann2004}. 
The inertness of the field
$\phi$ and the spatial homogeneity of its (gauge-invariant) modulus 
derives from the fact that it is obtained as a spatial average over 
nonpropagating, stable BPS saturated (zero energy-stress) field
configurations\footnote{Inertness of $\phi$ follows from the fact that a spatial 
average over nonpropagating fields can not generate propagating modes in
the effective field $\phi$. Thus an inhomogeneity of $|\phi|$ would
explicitely break the homogeneity of one-point functions which is in
contradiction to infinite-volume continuum thermodynamics.}. This can 
be checked explicitely by computing the curvature of $\phi$'s potential and by
comparing it with the squares of maximal resolution $|\phi|$ in the effective
theory and $T$ \cite{Hofmann2005}. So the effect of performing the spatial
coarse graining over Harrington-Shepard (anti)calorons is to make explicit the
spatial homogeneity and isotropy of the thermal system as contributed to 
by the only topologically nontrivial gauge-field
configurations which are admissible in computing a useful a priori estimate of the thermal ground state. The average effect of domainizations of 
the field $\phi$, which necessarily would lead to the
emergence\footnote{Microscopically, the emergence of these charges is
  understood by the dissociation of large-holonomy, that is,
  strongly-deformed-away-from-the-Harrington-Shepard-situation 
(anti)calorons.} of spatially localized and isolated magnetic charges \cite{Kibble}, is accounted for by the quantum field theoretic method of loop
expansion \cite{HofmannLE2006,SHG2006,KH2007,KHG2007}. 
Again, to each order of this expansion, spatial isotropy
and homogeneity are granted features. The sector with $Q=0$ and its
coupling to the inert field $\phi$ after spatial coarse-graining 
is uniquely determined by perturbative 
renormalizability \cite{tHooftVeltman}, gauge invariance, and the
spacetime symmetries of the underlying theory, for an explicit
presentation see Sec.\,\ref{Uni}. 

Apart from 1-PI reducible diagrams for the polarization tensor, 
which as compared to the tree-level situation can be resummed into mildly modified
dispersion laws, the expansion
into irreducible loop diagrams is expected to terminate at a finite loop order \cite{HofmannLE2006}. Numerically, there is a large hierarchy
between leading, next-to-leading, and next-to-next-to-leading 
corrections to the pressure \cite{SHG2006,KH2007}. As far as propagating,
effective gauge modes are concerned, this is the reason
why essentially all physics is contained in the 
tree-level quasi-particle spectrum and the resummed, 
one-loop polarization \cite{SHG2006,KH2007,LH2008}. 

Interactions between 
two topologically distinct 
sectors are hard to grasp when working with fundamental fields. As it 
turns out, in a situation determined by a global temperature scale $T$
(infinite-volume thermodynamics) the explicit 
consideration of fundamental fields and their interactions 
would necessitate external probes whose resolution needed to exceed the 
value $|\phi|$ emerging in terms of the Yang-Mills scale $\Lambda$ and $T$
when (sufficiently locally) integrating out the sector with $|Q|=1$.
That is, an attempt to saturate the partition function in terms of
fundamental field 
configurations, resolved with a higher momentum transfer than $|\phi|$,
necessarily introduces unsurmountable complications in assuring 
that the according approximations sustain 
homogeneous and isotropic thermalization.  

The main purpose of the present work is to 
investigate the physics of screened magnetic 
charges, being held responsible for the emergence of an area 
law for the spatial Wilson loop at high temperatures in lattice gauge
theory. We perform the analysis both in a hypothetic microscopic fashion and by directly appealing to 
the effective theory. As a warm-up to the effective theory for the deconfining 
phase we start by showing the uniqueness of the effective action in
Sec.\,\ref{Uni}. In Sec.\,\ref{KA} we review the argument given by 
Korthals-Altes for the existence of a spatial string tension as
orginated by isolated, screened and statistically independent 
magnetic charges. This argument assumes (anti)monopoles to behave like 
classical (resolvable) particles which, due to sufficiently strong
screening, separately 
contribute their magnetic flux through the minimal surface spanned by
the spatial contour. In Sec.\,\ref{MP} we use a two-loop correction
to the pressure, calculated in the effective theory and surviving the
high-temperature limit, to extract the average mass of the
screened and stable monopole-antimonopole system. Subsequently, their
hypothetic\footnote{By hypothetic we mean a physical number 
density seen by applying a sufficiently large spatial 
resolution $a^{-1}$ to typical configurations thermalized 
according to an associated perfect lattice action. The latter is
constructed such that 
the full partition function of the theory is invariant under a change
$a\to a^\prime$. Due to an efficient magnetic-charge neutralization 
it will turn out that this number density appears to be 
zero when applying the low resolution $|\phi|$ 
of our effective theory. The practical advantage of the 
low resolution $|\phi|$ is that it renders the perfect action to be
extremely simple and thus useful: The assumptions of \ref{KH} about screened 
magnetic monopoles and antimonopoles contributing their magnetic flux essentially uncancelled 
through the minimal contour spanning the spatial loop can be quantitatively 
checked by comparing screening length with mean spatial separation.} number density
$n$ is determined. As an aside, 
we demonstrate that, despite the hypothetic screening length being about three times
larger than the intermonopole distance in an {\sl ensemble} of, isolated, stable and screened
(anti)monopoles there is no large effect on their mass when comparing
with the BPS 
mass of an {\sl isolated} monopole-antimonopole pair immersed into a sea of
instable dipoles. The hypothetic number density $n$ in turn would imply a spatial string tension if the potentials of these 
stable magnetic (anti)monopoles would not overlap such as to cancel their
magnetic flux. Recall that the 
derivation of an area law, as performed 
in \cite{Korthals-Altes}, relies on a counting argument for
Poisson distributed (anti)monopoles. This argument assumes 
the independence of individual (anti)monopoles contributing to the total 
magnetic flux through a given spatial surface.  This independence, however, turns out to be strongly violated 
because stable magnetic objects essentially cancel their fluxes locally through their 
overlaping magnetic potentials. The immediate implication then is
that the spatial Wilson 
loop cannot exhibit area-law behavior at high temperatures. We confirm this result 
in Sec.\,\ref{SWLE} where 
the spatial Wilson loop is computed involving the lowest
nontrivial radiative corrections for the dispersion law of the massless mode 
in the effective theory: A
resummation of the one-loop polarization of the massless mode both
on- and off-shell. In
computing the spatial Wilson loop in real time 
we take into account the quantum and the thermal parts of 
the modified propagator for the massless mode, and we discuss the
potential contribution of the massive modes. We also 
consider the magnetic screening mass, defined as the zero-momentum 
limit of the square root of the screening function when setting the
temporal component of the momentum equal to zero
(Secs.\,\ref{screeningmass} and \ref{vacuumpart}). In Sec.\,\ref{sum} we summarize our
work and conclude.  

\section{Only minimal coupling of $Q=0$ with $|Q|=1$ sector in the effective action for the deconfining phase\label{Uni}}

Here we would like to argue that the effective action of the deconfining phase  
\eqb
\label{effdec}
S_{\tiny\mbox{dec}}=\mbox{tr}\,\int d^4x\,\left\{\frac12\,
  G_{\mu\nu}G_{\mu\nu}+D_\mu\phi D_\mu\phi+\frac{\Lambda^6}{\phi^2}\right\}\,.
\eqe
is complete. Namely, no operators representing local vertices of the field $a_\mu$ with three or 
more external legs together with (powers of) the field $\phi$ are possible while the nonexistence of operators of mass 
dimension larger than four {\sl solely} involving the gauge field $a_\mu$ is excluded 
by perturbative renormalizability \cite{tHooftVeltman}. Notice that nonlocal contributions 
to the effective action always can be expanded in powers of covariant derivatives. 

\begin{figure}
\begin{center}
\leavevmode
\leavevmode
\vspace{4.9cm}
\includegraphics{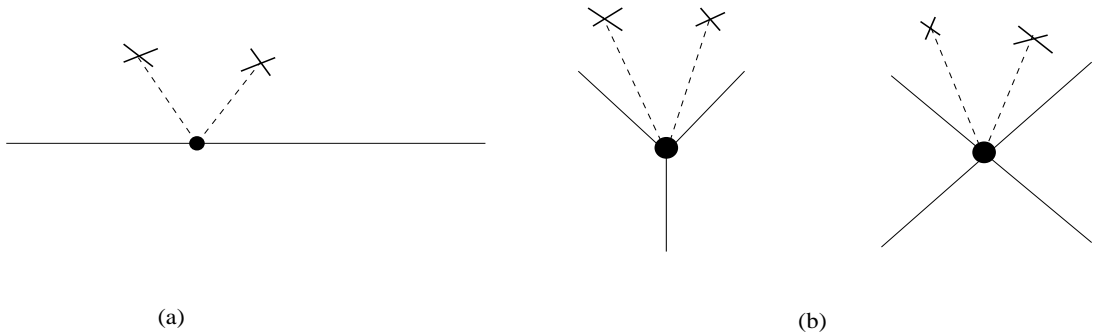}
\end{center}
\caption{\protect{\label{Fig-0A}} Allowed vertex (a) and examples of lowest mass dimension for excluded 
vertices in the effective theory (b). Solid lines represent the topologically trivial gauge field $a_\mu$ while a dashed 
line terminating in a cross corresponds to a local insertion of the operator $\phi$. }      
\end{figure}

Examples for excluded operators, still allowed by the spacetime symmetries and gauge invariance, would be 
\eqb
\label{excops}
\mbox{tr}\,\frac{1}{M^{2(3n-2)}}\left(G_{\mu\nu}[D_\mu\phi,D_\nu\phi]\right)^n\,,\ \ (n\ge 1)\,,
\eqe
$M$ representing some mass scale, while the operator $\mbox{tr}\,D_\mu\phi D_\mu\phi$ appearing in the action of 
Eq.\,(\ref{effdec}) is allowed. Fig.\,\ref{Fig-0A} depicts the situation of the allowed operator 
$\mbox{tr}\,D_\mu\phi D_\mu\phi$ and the $n=1$ operator of (\ref{excops}), which suffices to state our argument 
diagrammatically. 
Field $\phi$ is energy and
pressure free. (Its existence is owed to $|Q|=1$ (anti)selfdual fundamental
configurations of trivial holonomy.) Assuming a local vertex of $\phi$ 
with more than three external legs (in consistency with gauge
invariance and spacetime symmetries) coming from the gauge fields of the
coarse-grained trivial-topology sector $a_\mu$ necessarily would imply momentum
transfer to the field $\phi$, see (b) of Fig.\,\ref{Fig-0A} for a lowest-dimension example. That is, the probability 
of not transferring momentum to $\phi$ for nonvanishing momenta associated with the external legs 
of field $a_\mu$ is zero since this situation would correspond to a hypersurface in the space of all possible momentum transfers.  But a transfer of energy-momentum to $\phi$ would change its own
energy-momentum to nonvanishing values. This, however,
contradicts $\phi$'s BPS property. For the vertex of $\phi$ with
two external legs of the field $a_\mu$ (mass term in unitary gauge), as it is represented by
the term tr\,$(D_\mu\phi)^2$ in the effective
action of Eq.\,(\ref{effdec}) and depicted in (a) of Fig.\,\ref{Fig-0A}, no momentum is transferred. (Massiveness of the
off-Cartan topologically trivial gauge field $a_\mu$  emerges only after an
infinite resummation of such a vertex.) The fact that the ground state as
such has a finite energy density in the effective theory is owed to a pure
gauge configuration. This configuration on the effective level sums up interactions
(energy-momentum transfers {\sl larger} than $|\phi|$, that is at distances smaller than $|\phi|^{-1}$) of topologically
trivial fundamental gauge fields with (anti)calorons which are not visible
at resolution lower than $|\phi|$ but lift the ground-state energy density from zero to a finite
value (temporary shift of holonomy plus radiative corrections).

To conclude, the existence of vertices involving $\phi$ and more than
two legs of the coarse-grained topologically trivial gauge field $a_\mu$ in the effective theory for the deconfining phase 
is excluded by the BPS property of $\phi$: The very existence of $\phi$,
which is guaranteed to emerge from the $|Q|=1$ sector of the theory upon
spatial coarse-graining \cite{HerbstHofmann2004,Hofmann2005,Hofmann2007}, 
would be contradicted by it absorbing energy-momentum {\sl after} coarse-graining.

As for the preconfining phase, renormalizability is trivial since we
coarse-grain over the noninteracting $Q=0$ sector of a U(1) (or U(1)$^2$
in the case SU(3)) gauge 
theory. Thus the same arguments as for the deconfining phase when 
adjusted to this simpler, abelian situation do fix the action of the
associated Higgs model \cite{Hofmann2005} uniquely in the preconfining phase.

\section{Korthals-Altes' derivation of fundamental spatial string
  tension\label{KA}}

We only consider the case of deconfining SU(2) Yang-Mills thermodynamics
with Yang-Mills scale $\Lambda$. In
the effective theory for this phase the tree-level situation is described by an
inert adjoint and spatially homogeneous scalar field of modulus
$|\phi|=\sqrt{\frac{\Lambda^3}{2\pi T^3}}$, 
a pure-gauge configuration of the coarse-grained $Q=0$ sector, 
and propagating gauge modes (one direction of the
SU(2) algebra massless, two directions of mass $2e|\phi|$). This
effective theory neatly decomposes physics into 
a thermal ground state (equation of state equal to that of a cosmological constant
but with a temperature dependent energy density) and thermally fluctuating
(quasi)particles \footnote{Due to the existence of the maximal resolution
  scale $|\phi|$ the effect of having coarse-grained gauge modes propagate off
  their mass shell without explicit interaction is, on the one-loop level, completely negligible for
  thermodynamical quantities such as the pressure \cite{Hofmann2005}.}. 
Notice that on tree-level this `a-priori-estimate' for the thermal ground state only
captures the physics of small-holonomy (anti)calorons which are associated
with short-lived monopole-antimonopole pairs. This leads to an 
explicit manifestation of spatial homogeneity and isotropy in terms of $\phi$
and a pure gauge $a^{gs}_\mu$ as it is demanded
by thermodynamics. The effects attributed to the 
rare occurrence of isolated, long-lived, and screened
magnetic charges in the plasma are captured by certain radiative corrections. This
can be pictured as an implicit average over domainized $\phi$-field 
configurations \cite{Kibble} of typical domain size smaller than $|\phi|^{-1}$
facilitated by the quantum field theoretic 
method of loop expansion in effective variables. Notice that spatial
homogeneity and isotropy are explicitely honored at each loop order.   

Before we actually establish the connection between radiative corrections to
the pressure as computed in the effective theory and the hypothetic 
physics of screened magnetic 
monopoles we would like to review the arguments given by Korthals-Altes
\cite{KH} 
on how a spatial string tension emerges once the existence and
independence of
isolated and screened (magnetic screening length $l_s$) magnetic charges is assumed in the
plasma. Korthals-Altes considers a quadratic, spatial contour of 
side-length $L$ and a number density of monopoles $n_M$. Since monopoles are
created pairwise by the dissociation of large-holonomy calorons
\footnote{Notice that the occurrence of a nontrivial caloron holonomy in the sense
of a spacetime independent parameter, as it occurs in (anti)selfdual Yang-Mills 
fields, is unphysical because of the vanishing impact of these configurations
onto the partition function \cite{GPY1981}. In case of small
holonomy it is still
useful though to think of the holonomy parameter as possessing a
finite-support real-time dependence of width that is comparable to the
life-time of the monopole-antimonopole pair. Upon a continuation of this real-time
dependence back to the euclidean a departure from
(anti)selfduality takes place.} one has $n_M=n_A\equiv n$ where $n_A$ is the
number density of antimonopoles. Furthermore, he assumes (and justifies this
assumption by lattice data obtained with the Wilson action) 
that monopoles and
antimonopoles fluctuate in an independent way and that they are rare. This is
certainly true if the screening length $l_s$ is much smaller than the mean
interparticle distance $d=\left(\frac{1}{n}\right)^{1/3}$. We will show
in Sec.\,\ref{dens}, however, that this situation is actually {\sl not}
realized. Still, the mass of the screened
monopole-antimonopole system 
is close to its BPS bound. 

By virtue of the length scale $l_s$ Korthals-Altes considers 
a slab of thickness $2l_s$ containing magnetic quasiparticles. These quasiparticles contribute a mean 
magnetic flux through the minimal surface spanned by the afore-mentioned 
contour. Only
those quasiparticle that are contained within the slab are held responsible
for the flux. A more refined treatment introducing no such
constraint but taking into account
a Yukawa-like potential for the static magnetic field sourced by each of these
objects 
yields similar results. 

As a brief interlude, we name an important property of these 
monopole-antimonopole pairs. Since we know now that pairs of magnetic monopoles and
antimonopoles are liberated through the dissociation of large-holonomy
(anti)calorons we also know -- by studying the BPS (anti)monopole constituents in
such (anti)calorons \cite{NahmVanBaalLeeLu} -- 
that the combined mass $m\equiv m_M+m_A$ of the monopole and its
antimonopole (a holonomy-independent quantity) is roughly given as 
\cite{NahmVanBaalLeeLu}
\eqb
\label{monAmonMass}   
m\sim\frac{8\pi^2 T}{e}=\sqrt{8}\pi\,T\sim 8.89\,T\,.
\eqe
Here the high-temperature plateau-value $e=\sqrt{8}\,\pi$ 
for the {\sl effective} gauge coupling $e$ was used 
\cite{Hofmann2005,Hofmann2007,GiacosaHofmann2007}. 
Eq.\,(\ref{monAmonMass}) is valid for an isolated, noninteracting (test)
monopole-antimonopole system with their magnetic charges being 
screened by the surrounding, short-lived magnetic dipoles belonging to
small-holonomy (anti)calorons. This is the situation described by the
one-loop expressions for thermodynamical quantities in the effective theory. 
By virtue of Eq.\,(\ref{monAmonMass}) magnetic monopoles and
antimonopoles appear as nonrelativistic objects in the plasma. 
They are Boltzmann suppressed as
\eqb
\label{Boltzmann}
P_{M+A}\sim \exp(-\sqrt{8}\pi)\sim 1.4\times 10^{-4}\,,
\eqe
and thus occur rarely as compared to the instable pairs 
that are associated with small (anti)caloron holonomies.   

Now Korthals-Altes exponentiates the 
normalized flux $\Phi_{l=1}$ of a single (anti)monopole through the minimal
surface spanned by the contour in the limit $L\to\infty$ as
\eqb
\label{V(L)}
\left.V(L)\right|_{l=1}\equiv \exp\left(2\pi i \frac{\Phi_{l=1}}{Q}\right)\,,
\eqe
where $Q$ is the magnetic charge. 
Depending on the sign of $Q$ and on the location w.r.t. the minimal 
surface one has $\Phi_{l=1}=\pm \frac12$ and thus
\eqb
\label{V(L)aus}
\left.V(L)\right|_{l=1}\equiv -1\,.
\eqe
Korthals-Altes now assumes 
the probability $P(l)$ for the occurrence of $l$
charges within the slab to be given by the Poisson distribution 
\eqb
\label{Poisson}
P(l)=\frac{\bar{l}^l}{l!}\,\exp(-\bar{l})\,,
\eqe
where $\bar{l}$ is the mean value of the number of magnetic charges 
contained inside the slab. One then obtains 
\eqb
\label{meanflux}
\bar{V}(L)=\sum_{l=0}^\infty P(l) (-1)^l=\exp(-2\bar{l})\,.
\eqe
Since $\bar{l}=4\,A(L)l_s n(T)$, where $A(L)$ is the minimal surface and
$n(T)$ is the (temperature-dependent) density of monopoles (or antimonopoles
or monopole-antimonopole pairs), we observe that the Wilson loop shows area
law with tension $\sigma_s$ given as 
\eqb
\label{sigma_s}
\sigma_s=8\,l_s n(T)\,.
\eqe
Notice that 
the hypothetic magnetic screening length $l_s$ is given in terms of $n$
and $T$ as \cite{KH}
\eqb
\label{magscl}
l_s=\frac{1}{g}\sqrt{\frac{T}{n}}\,,
\eqe
where $g$ is the magnetic coupling: $g\equiv\frac{4\pi}{e}$.

\section{Monopole physics from two-loop correction to the pressure \label{MP}}

In this section we would like to investigate the high-temperature
implications of the radiative corrections to the pressure, calculated in
the effective theory, for the physics of unresolved, stable and screened
magnetic monopoles and antimonopoles as they emerge from the
dissociation of large-holonomy (anti)calorons. Notice 
that the density of these objects is inferred purely 
energetically without making reference to the existence or nonexistence 
of their net magnetic fluxes.

\subsection{Mass of interacting monopole-antimonopole system}   

On distances larger than the minimal spatial length \cite{Hofmann2005}
\eqb
\label{phiinTc}
|\phi|^{-1}=\frac{(13.87)^{3/2}}{2\pi T_c}\,\sqrt{\frac{T}{T_c}}
\eqe
and not taking into account radiatively liberated stable and isolated
monopole-antimonopole pairs, the depletion of a single magnetic test charge is
described by the effective, electric coupling $e=\sqrt{8}\pi$ for $T\gg
T_c$. Notice that the (constant) value of $e$ signals that for $T\gg T_c$ the
system forgets about the presence of the Yang-Mills scale 
$\Lambda$ as far as its propagating degrees of freedom are concerned: The 
thermodynamics of these modes then solely is determined by 
topology and temperature. This fact may be conceived as a
nonperturbative manifestation of asymptotic freedom.

It is important to stress that the effective, electric coupling rapidly 
approaching the constant $e=\sqrt{8}\pi$ at increasing temperature is a 
consequence of an evolution equation assuring that the interaction of the thermal ground state with propagating quasiparticle excitations honours Legendre transformations 
\cite{Hofmann2005, Hofmann2007}. The constancy $e=\sqrt{8}\pi$ is approached 
power-like fast in $T$. So the error of using the constant 
$e=\sqrt{8}\pi$, which is a lower bound on the behavior close to the phase transition, at moderate temperatures dies off in a power-like way. In \cite{SHG2006} 
we have used the exact evolution of $e$ and, indeed, have observed, within a few percent variation, the asymptotic constancy of the two-loop correction to the pressure divided by $T^4$ starting from $T=3\,T_c$. Asymptotic constancy of $e$ is not to be confused with the behavior of the {\sl fundamental} gauge coupling which can be defined via the trace anomaly of the energy-momentum tensor \cite{GiacosaHofmann2007}. The latter approaches zero logarithmically 
slowly with increasing $T$.

The holonomy-independent sum $m$ of the masses of the BPS monopole and 
BPS antimonopole (BPS mass), being the constituents of a nontrivial-holonomy (anti)caloron, is given
as \cite{Hofmann2005,NahmVanBaalLeeLu}
\eqb
\label{sumofmasses}
m_{>|\phi|^{-1}}=\frac{8\pi^2 T}{e}=\sqrt{8}\pi\,T\,.
\eqe
Notice that Eq.\,(\ref{sumofmasses}) describes the situation of a pair of BPS 
saturated monopole and antimonopole placed as test charges into a surrounding where
small-holonomy (anti)calorons generate short-lived magnetic 
dipoles effectively leading to a finite renormalization of the 
magnetic charge of the test particles. Here no departure from the BPS 
limit is implied, and the U(1) gauge 
field of the test (anti)monopole still is infinite-range. A linear
superposition of these potentials then leads to a dipole form which for
large distances $R$ decays like $1/R^3$. This
corresponds to the approximation of a massless and 
two massive {\sl free} thermal quasiparticles in the
effective theory (tree-level).    
 
On the level of radiative corrections, however, we are 
concerned with the physics of isolated, stable magnetic charges whose average 
distance $n^{-1/3}$ at high temperature will be given by $n^{-1/3}=c/T$
where $c$ is a positive, real constant (determined in
Sec.\,\ref{dens}). Thus 
\eqb
\label{ratiolength}
|\phi|\,n^{-1/3}=2\pi c\,\left(\frac{T_c}{13.87\, T}\right)^{3/2}\,,
\eqe
which is smaller than unity for sufficiently high $T$. In fact, an 
estimate implies that for $T\ge 1.91\,T_c$ isolated and screened
(anti)monopoles are not resolved in the effective theory. 
This estimate uses the high-temperature value
$e=\sqrt{8}\,\pi$ for the effective gauge coupling. A more careful 
investigation for $T$ shortly above $T_c$ shows that isolated and screened
(anti)monopoles actually are never fully resolved in the effective theory for the 
deconfining phase. However, their effect on the propagation of the tree-level
massless gauge mode is sizable at temperatures a few times $T_c$ \cite{SHG2006,LH2008}.  

To make contact with degrees of freedom, whose
collective long-distance effects are detectable (antiscreening and screening of effective 
gauge-field propagation) but which never appear 
explicitely in the effective theory, we may consider the situation of thermally fluctuating,  
free quasiparticles (one-loop truncation of loop expansion of the
pressure) 
as the asymptotic starting 
point. Small interactions between the quasiparticles, as they are described by
radiative corrections (for $T\gg T_c$ only the two-loop diagram 
involving a four-vertex connecting on shell a massless with
either of the two massive modes survives) introduce interesting
physics but do not change the energy density of the 
one-loop situation. The important implication then is that radiative
corrections to the one-loop situation, calculable in the effective theory,
must be {\sl cancelled} by the physics of unresolved degrees of freedom on the
level of thermodynamic quantities. By
the result of \cite{Diakonov}, which shows that (anti)calorons dissociate upon
strong quantum deformation into isolated magnetic monopole--antimonopole pairs, we are led
to conclude that these degrees of freedom are indeed isolated, stable, and screened magnetic
monopoles and antimonopoles.   

Thus we need to compute the mass $m$ of monopole-antimonopole systems for distances of the order of 
$n^{-1/3}$. Since the spatial coarse-graining, which determines the field
$\phi$, saturates exponentially fast on distances of a few $\beta\equiv 1/T$
\cite{Hofmann2007,HerbstHofmann2004} we expect that 
\eqb
\label{ratiomasses}
\frac{m}{m_{>|\phi|^{-1}}}=O(1)  
\eqe
which, indeed, is the case. Let us show this. 

As mentioned above, at large temperatures the two-loop diagram for the pressure 
involving a four-vertex connecting a massless with
either of the two massive modes is the only surviving 
radiative correction \cite{SHG2006,HerbstHofmannRohrer2004}. For $T\gg T_c$
one has \cite{SHG2006}
\eqb
\label{Prescorr} 
\frac{\Delta P_{\tiny\mbox{2-loop}}}{P_{\tiny\mbox{1-loop}}}=-4.39\times
10^{-4}\,,
\eqe
where $\Delta P_{\tiny\mbox{2-loop}}\equiv
P_{\tiny\mbox{2-loop}}-P_{\tiny\mbox{1-loop}}$. 
Since for $T\gg T_c$ we have \cite{Hofmann2005}
\eqb
\label{Pres1loop}
P_{\tiny\mbox{1-loop}}=\frac{4}{45}\pi^2\,T^4
\eqe
the correction $\Delta P_{\tiny\mbox{2-loop}}$ is also proportional to
$T^4$. Now $\Delta P_{\tiny\mbox{2-loop}}$ and the associated energy density
$\Delta\rho_{\tiny\mbox{2-loop}}$, which both are negative, must obey
the following relation\footnote{Legendre transformations are linear and thus 
hold for each order in the loop expansion separately when integrating out
residual quantum fluctuations.}   
\eqb
\label{enecorr}
\Delta\rho_{\tiny\mbox{2-loop}}=T\frac{d\Delta
  P_{\tiny\mbox{2-loop}}}{dT}-\Delta P_{\tiny\mbox{2-loop}}\,.
\eqe
Eq.\,(\ref{enecorr}) and the fact that $\Delta
P_{\tiny\mbox{2-loop}}\propto T^4$ imply an equation of state 
\eqb
\label{eos}
\Delta\rho_{\tiny\mbox{2-loop}}=3\,\Delta P_{\tiny\mbox{2-loop}}\,,
\eqe
and Eq.\,(\ref{Prescorr}) tells us that the thermal 
energy density $\Delta\rho_{M+A}$, attributed to the
presence of unresolved, screened, isolated pairs of (no
longer BPS saturated) magnetic monopoles and 
antimonopoles is given as\footnote{Since the process of (anti)caloron dissociation, creating pairs of isolated and screened magnetic 
(anti)monopoles subject to an exact, overall charge neutrality, is irreversible the density of these objects is sharply fixed 
at a given temperature. Thus there is no chemical potential associated with monopole-antimonopole pairs.}
\eab
\label{monpairdensenrg}
 \frac{\Delta\rho_{M+A}}{T^4}&\equiv&\frac{1}{2\pi^2} \int_{\mu}^\infty
 dy\,\sqrt{y^2-\mu^2}\,y^2\,n_B(y)\stackrel{!}=-\frac{\Delta\rho_{\tiny\mbox{2-loop}}}{T^4}\nonumber\\
 &=&4.39\times 10^{-4}\times \frac{4}{15}\pi^2\,,
\eae
where $\mu\equiv m/T$ and $n_B(y)\equiv 1/(\exp(y)-1)$. The only
positive solution $\mu$ of Eq.\,(\ref{monpairdensenrg}) is
numerically given as $\mu=10.1224$. Thus we have 
\eqb
\label{ratiomassesnum}
\frac{m}{m_{>|\phi|^{-1}}}=1.139\,.  
\eqe
The effect of all other, screened, isolated, and stable
(anti)monopoles, generated by the dissociation of large-holonomy
(anti)calorons, hence is a lift of the mass of the system by about 
14\% above the BPS bound even though the hypothetic magnetic
screening length $l_s$ at high $T$ is larger than the mean 
interparticle distance $\bar{d}$, see Sec.\,\ref{dens}. 

\subsection{Monopole density, (anti)monopole distance, and screening length\label{dens}}

Knowing that $\mu=10.1224$ we now can 
compute the hypothetic monopole density $n$ as
\eqb
\label{density}
n=\frac{T^3}{2\pi^2}\int_\mu^\infty
dy\,y\sqrt{y^2-\mu^2}\,n_B(y)=9.799\times 10^{-5}\,T^3\,.
\eqe
For reasons of symmetry there is a {\sl universal} mean monopole-antimonopole
distance $\bar{d}$. That is, the distance between a monopole and its
antimonopole, both stemming from the dissociation of the same
(anti)caloron, is, on average, the same as the distance between a monopole
adjacent to an antimonopole who did not originate from the same (anti)caloron. 
From Eq.\,(\ref{density}) we have
\eqb
\label{distance}
\bar{d}=n^{-1/3}=21.691\,\beta\,.
\eqe
Thus the constant $c$ in Eq.\,(\ref{ratiolength}) is given as
$c=21.691$, and the right-hand side of (\ref{ratiolength}) is smaller
than unity for $T>2\,T_c$. That is, isolated and screened
(anti)monopoles are {\sl not resolved} at high temperatures in the
effective theory. Because the magnetic flux of a given pair essentially
cancels due to the large overlap of magnetic potentials no area law is 
to take place for the spatial Wilson
loop, see Eq.\,(\ref{ratioscdis}). 
This is in accord with our investigation performed in Sec.\,\ref{SWLE}. 

By virtue of Eqs.\,(\ref{magscl}) and (\ref{density}) the hypothetic magnetic 
screening length $l_s$ is given as
\eqb
\label{magscl}
l_s=\frac{1}{g}\sqrt{\frac{T}{n}}=\frac{e}{4\pi}\,(21.691)^{3/2}\,\beta=71.43\,\beta\,.
\eqe
Thus we have 
\eqb
\label{ratioscdis}
\frac{l_s}{\bar{d}}=3.293\,.
\eqe
Eq.\,(\ref{ratioscdis}) tells us that monopoles and 
antimonopoles are not far separated on the scale of their screening 
length. Notice that both of these length scales are much 
smaller than $|\phi|^{-1}$. However, the impact of all other stable and screened 
monopoles and antimonopoles on a given stable 
pair is small due to efficient cancellations of mutual 
attraction or repulsion. This is expressed by the small lift of the BPS
mass $m_{>|\phi|^{-1}}$, compare with Eq.\,(\ref{ratiomassesnum}).

\section{Spatial Wilson loop in effective variables\label{SWLE}}

The effective theory for deconfining SU(2) Yang-Mills theory and the
computation of radiative corrections to the pressure 
is described at length in \cite{Hofmann2005,Hofmann2007,SHG2006}. We only
name those results explicitely that are directly needed and for the
remainder refer the reader to these 
papers.    

\subsection{Magnetic screening mass\label{screeningmass}}

The SU(2) gauge symmetry of the fundamental action is 
dynamically broken to U(1) by coarse-grained (anti)calorons. Notice that the order parameter 
$\phi$ of this gauge-symmetry breaking is determined in a highly 
nonlocal way \cite{Hofmann2007,HerbstHofmann2004} thus appealing to the strong magnetic-magnetic 
correlations mediated by the $|Q|=1$ (anti)caloron configuration of trivial 
holonomy \cite{HS1977}.

In unitary-Coulomb gauge there are in the effective theory on 
tree-level one massless vector excitation (hencefore referred to as $\gamma$)
and two thermal vector quasiparticle excitations (hencefore referred to as
$V^\pm$). The transversal $\gamma$ polarization
tensor $\Pi^{\mu\nu}$ is decomposed as 
\eqb
\label{Pidec}
\Pi^{\mu\nu}=G(p_0,\vec{p})\,P^{\mu\nu}_T+F(p_0,\vec{p})\,P^{\mu\nu}_L
\eqe
where 
\eqb
\label{PL}
P^{\mu\nu}_L\equiv \frac{p^\mu p^\nu}{p^2}-g^{\mu\nu}-P^{\mu\nu}_T\,,
\eqe
and
\eab
\label{transproj}
P^{00}_T&=&P^{0i}_T=P^{i0}_T=0\,,\nonumber\\ 
P^{00}_T&=&\delta^{ij}-p^ip^j/\vec{p}^2\,.
\eae
The functions $G(p_0,\vec{p})$ and $F(p_0,\vec{p})$ determine the propagation of 
the interacting $\gamma$ mode. For $\mu=\nu=0$ Eq.\,(\ref{Pidec}) yields upon rotation 
to real-time 
\eqb
\label{P00F}
F(p_0,\vec{p})=\left(1+\frac{p_0^2}{p^2}\right)^{-1}\,\Pi^{00}\,.
\eqe
The function $F(p_0,\vec{p})$ measures the screening of electric 
fields which we are not interested in when discussing the {\sl spatial} Wilson
loop. For propagation of the $\gamma$ mode into the 3-direction we have
\eqb
\label{gxxgyy}
\Pi_{11}=\Pi_{22}=G(p_0,\vec{p})\,.
\eqe
From the magnetic screening function $G(p_0,\vec{p})$ we obtain a definition for the magnetic screening $m_s$ as
\eqb
\label{ms}
m_s\equiv \lim_{\vec{p}\to 0}\sqrt{\mbox{Re}\,G(p_0=0,\vec{p})}\,.
\eqe
In Eqs.\,(\ref{gxxgyy}) and (\ref{ms}) we have suppressed the dependence on temperature
of the magnetic screening function $G$. The radiatively generated mass scale $m_s$ measures the exponential
decay rate of an externally applied, homogeneous magnetic field
$B_i=\epsilon_{ijk} G^3_{jk}$ penetrating the
Yang-Mills plasma. In the effective theory, 
the screening function $G$ only has support 
for the four-momentum $p$ satisfying the constraint 
\eqb
\label{constr}
|p^2-G(p^0,\vec{p})|\le |\phi|^2\,,
\eqe
where $|\phi|$ is the (temperature-dependent) modulus of the 
inert, adjoint scalar field $\phi$ emerging as a consequence of
spatial coarse-graining over the stable BPS sector with topological
charge modulus $|Q|=1$. Eq.\,(\ref{constr}) is the condition that a propagating effective tree-level massless gauge mode in physical Coulomb-unitary gauge must not be further off its radiatively induced mass shell 
then the scale $|\phi|$ of maximal resolution.

\begin{figure}
\begin{center}
\leavevmode
\leavevmode
\vspace{4.9cm}
\includegraphics{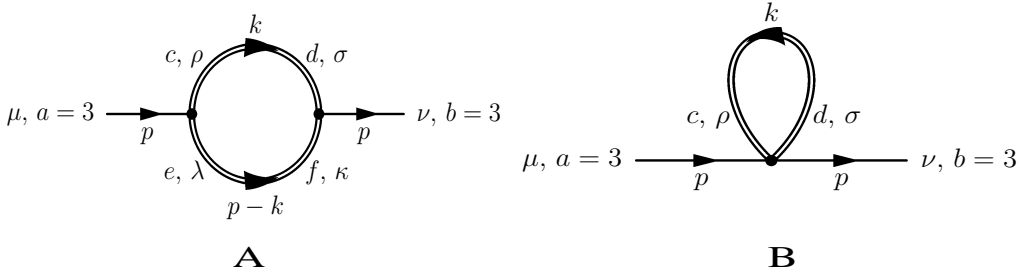}
\end{center}
\caption{\protect{\label{Fig-1}} The diagrams for $\gamma$'s 
polarization tensor $\Pi_{\mu\nu}$.}      
\end{figure}

In Fig.\,\ref{Fig-1} the two diagrams a priori 
contributing to $\Pi^{\mu\nu}$ are depicted. 
Using standard Feynamn rules, we have for diagram A 
\begin{equation}
\label{Anproc}
\begin{split}
\Pi^{\mu\nu}_{A}(p)\,=\,& 
\frac{1}{2i}\int\frac{d^4k}{(2\pi)^4} e^2 
	\epsilon_{ace}[g^{\mu\rho}(-p-k)^\lambda+g^{\rho\lambda}(k-p+k)^\mu+g^{\lambda\mu}(p-k+p)^\rho]\times\\
& \epsilon_{dbf}[g^{\sigma\nu}(-k-p)^\kappa+g^{\nu\kappa}(p+p-k)^\sigma+g^{\kappa\sigma}(-p+k+k)^\nu]\times\\
&(-\delta_{cd})\left(g_{\rho\sigma}-\frac{k_\rho k_\sigma}{m^2}\right) 
\left[\frac{i}{k^2-m^2}+2\pi\delta(k^2-m^2)\,n_B(|k_0|/T) \right]\times\\
&(-\delta_{ef})\left(g_{\lambda\kappa}-\frac{(p-k)_\lambda(p-k)_\kappa}{(p-k)^2}\right)\times\\
&\left[\frac{i}{(p-k)^2-m^2}+2\pi\delta((p-k)^2-m^2)\,n_B(|p_0-k_0|/T) \right]\,.
\end{split}
\end{equation}
From the one-loop evolution \cite{Hofmann2005} we know that $e\ge
\sqrt{8\pi}$ \cite{Hofmann2007}. Due to constraint $|k^2-m^2|\le
|\phi|^2$ \cite{Hofmann2005}, where $m=2e\,|\phi|$, the vacuum part in the $V^\pm$ propagator is 
forbidden. For an explicit presentation of the $\gamma$ and $V^\pm$ 
real-time propagators see Eqs.\,(\ref{Vpm}) and (\ref{gamma}).

Diagram B 
reads
\begin{equation}
\label{vactad}
\begin{split}
\Pi^{\mu\nu}_{B}(p)\,=\,&\frac{1}{i} \int \frac{d^4k}{(2\pi)^4} 
(-\delta_{ab}) \left( g_{\rho\sigma}-\frac{k_\rho k_\sigma}{m^2} \right)  
\left[\frac{i}{k^2-m^2}+2\pi\delta(k^2-m^2)n_B(|k_0|/T) \right]\times\\
 & \quad (-ie^2)[
 \epsilon_{abe}\epsilon_{cde}(g^{\mu\rho}g^{\nu\sigma}-g^{\mu\sigma}g^{\nu\rho})
 +\epsilon_{ace}\epsilon_{bde}(g^{\mu\nu}g^{\rho\sigma}-g^{\mu\sigma}g^{\nu\rho})+\\
 &\quad\epsilon_{ade}\epsilon_{bce}(g^{\mu\nu}g^{\rho\sigma}-g^{\mu\rho}g^{\nu\sigma})]\,.
\end{split}
\end{equation}
Again, the part in Eq.\,(\ref{vactad}) arising from the vacuum contribution in 
Eq.\,(\ref{Vpm}) vanishes. For the four-vertex in diagram B the
following constraint holds \cite{Hofmann2005}: $|(p+k)^2|\le
|\phi|^2$. 

\begin{figure}
\begin{center}
\vspace{5.8cm}
\includegraphics{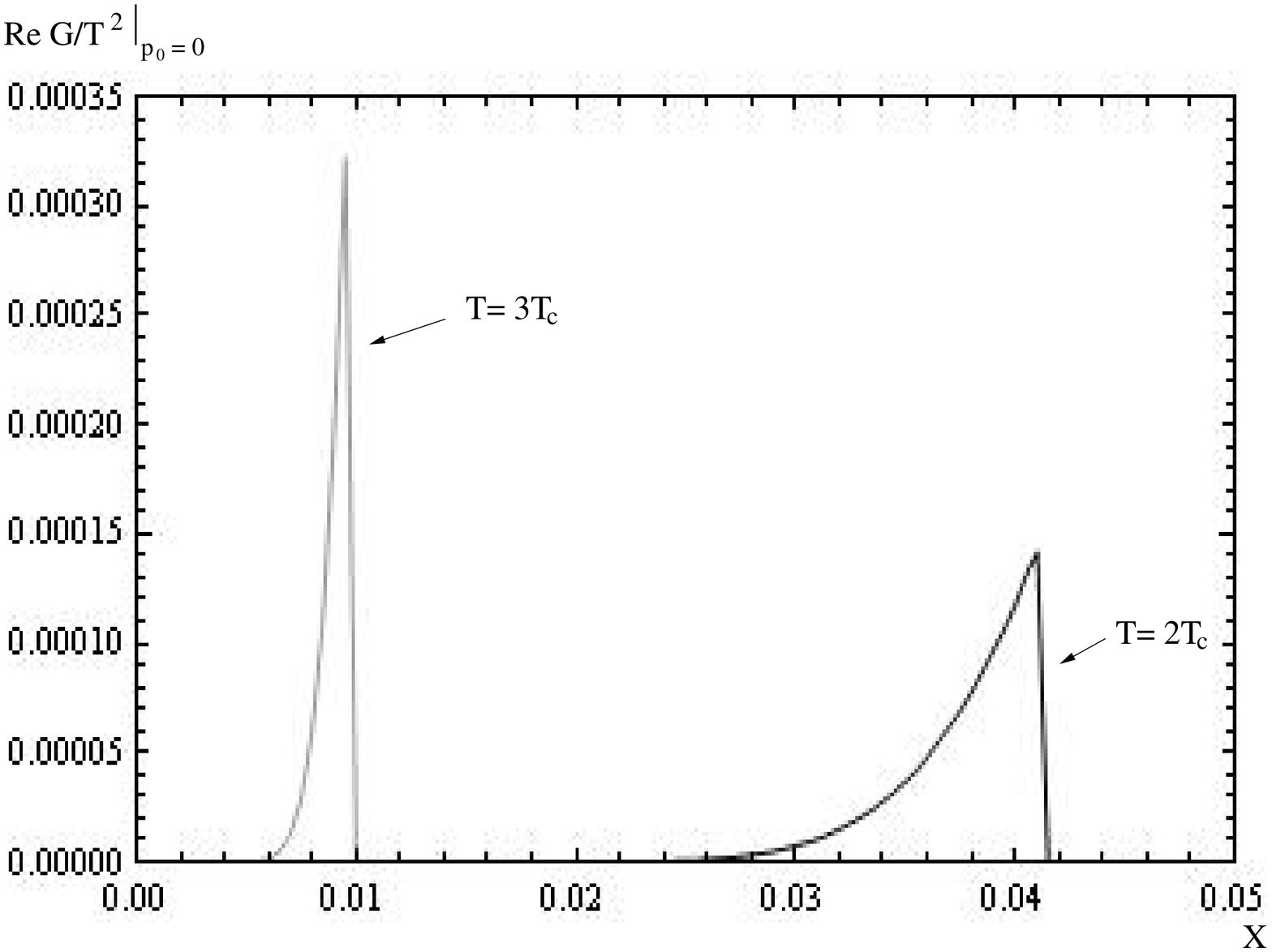}
\end{center}
\caption{\protect{\label{G}}The function $\mbox{Re}\,G/T^2$ in dependence of
  $X=|\vec{p}|/T$ setting $p_0=0$ for $T=2\,T_c$ (black) and $T=3\,T_c$ (gray).}
\end{figure}

It is easily seen that only diagram B contributes to 
$m_s$. Generalizing $m_s\to m_s(p_0=0,|\vec{p}|)$ within the finite
support in $\vec{p}$ given by the condition (\ref{constr}) now
specializing to $|\vec{p}^2+G(p^0=0,\vec{p})|\le |\phi|^2$, we obtain a behavior as
depicted in Fig.\,\ref{G} in dependence of
  $X=|\vec{p}|/T$. 
From Fig.\,\ref{G} and comparing with Eq.\,(\ref{ms}) one sees 
that $m_s=0$ which is in agreement with our
microscopic analysis of Sec.\,\ref{MP}. Recall that due to the
unresolvability of stable and screened monopole-antimopole 
pairs no net monopole or antimonopole density\footnote{In contrast to
  Sec.\,\ref{dens} we are here concerned with the monopole density 
as seen in our {\sl effective} theory at 
resolution $|\phi|$ and not with the 
hypothetic monopole density detected by a higher resolution.} 
is detected and thus, according 
to Eq.\,(\ref{magscl}), $l_s\equiv 1/m_s=\infty$.

\subsection{Generalities on spatial Wilson loop in effective theory\label{GSWE}}

To compute a Wilson loop in effective 
variables associated with a given, finite resolution $\mu$ is in general something different than the definition in fundamental variables at infinite resolution demands. 
However, at finite temperature in the deconfining phase of a 
Yang-Mills theory there is only one resolution scale where the effective action minimally differs from the fundamental action: at $\mu=|\phi|$ technically topologically trivial and nontrivial field configurations 
are separated modulo mass generation and constraints on the hardness of quantum fluctuations of the former imposed by the latter. Since an ensemble average over 
isolated and screened magnetic monopoles is generated by the radiative corrections 
in the effective theory it is suggestive that the spatial Wilson loop evaluated on propagating gauge mode conveying the magnetic flux of the effective theory, with a resummation of the lowest-order radiative effects (polarization tensor) indeed, measures the flux sourced by the ensemble of 
magnetic monopoles determining the polarization tensor. This argument is purely intuitive, 
and no proof of this suggested property is yet available.      

Here we quote some general results on the calculation 
of exponentiated one-effective-gauge-boson exchanges within the spatial
quadratic contour $C$ of side-length $L$. 

\begin{figure}
\begin{center}
\vspace{5.3cm}
\includegraphics{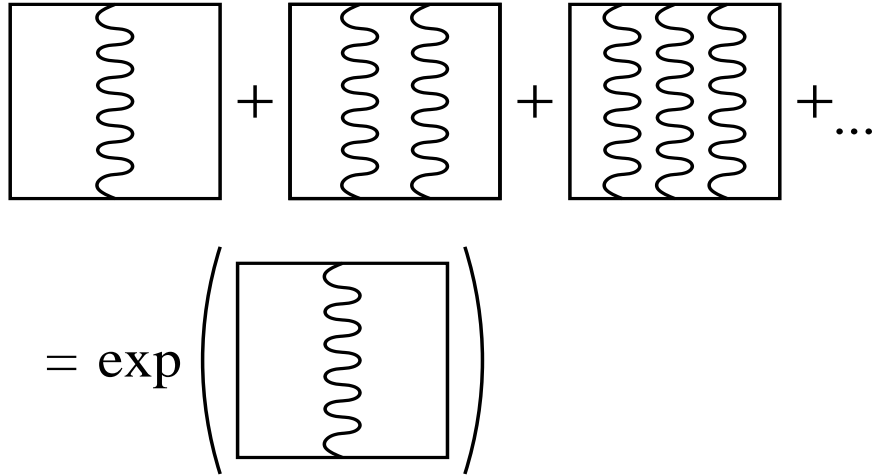}
\end{center}
\caption{\protect{\label{Wilson}Illustration of the diagrammatic
    approach. 
Every summand in this formula represents a whole 
class of $N$-gauge-boson exchange diagrams, which is encoded in terms of a
(suppressed) 
combinatoric factor.}}
\end{figure}

In order to calculate the spatial string tension, we use an expansion
into loops for the $N$-point functions in the effective theory \cite{Bassetto}. Because of the rapid
numerical convergence of the effective loop expansion
\cite{SHG2006,HofmannLE2006,KH2007} we are content here 
with a resummation of the one-loop polarization tensor for the $\gamma$ 
mode into its effective propagator and the tree-level propagator for the
$V^\pm$ modes (2-point functions). The exchange of these 
modes is subsequently exponentiated in order to take into 
account higher $N$-point functions in a trivial way, see Fig.\,\ref{Wilson}.   

Thus we can write for the logarithm of $W[C]$
\begin{equation}
\ln W[C]=-\frac{1}{2}C_F \oint\!dx_\mu dy_\nu\,D_{\mu\nu}(x-y)\,,
\end{equation}
where $C_F$ denotes the Dynkin index in the fundamental representation, 
defined as the normalization factor of the generators of the algebra (i.e. $C_F\equiv\frac{1}{2}$), and
\begin{equation}
D_{\mu\nu}=\sum\limits_{a=1}^3 D_{\mu\nu}^{(a)}
\end{equation}
is the sum of the tree-level propagators ($V^\pm$ corresponds to $a=1,2$) and the one-loop 
resummed propagator ($\gamma$ corresponds to $a=3$). 

The tree-level propagators for $a=1,2$ in 
position space are given as the Fourier transformations of their momentum space counterparts
\begin{equation}
\label{Vpm}
D_{\mu\nu}^{1,2}(\beta,x-y)
=-\int\!\frac{d^4p}{(2\pi)^4}\,\e^{-ip(x-y)}\left(g_{\mu\nu}-\frac{p_\mu p_\nu}{m^2}
\right)\left[\frac{i}{p^2-m^2+i\varepsilon}
+2\pi\delta(p^2-m^2)n_B(\beta|p_0|)\right]\,.
\end{equation}
The one-loop resummation of the magnetic part\footnote{We are not
  interested in electric screening effects since we study the {\sl
    spatial} Wilson loop.} of the polarization
tensor (screening function $G$) into the dressed 
$\gamma$ propagator, see
Eq.\,\ref{Pidec}, yields the following result 
\begin{equation}
\label{gamma}
D_{\mu\nu}^{3}(\beta,x-y)
=\int\!\frac{d^4p}{(2\pi)^4}\,\e^{-ip(x-y)}
\left[P^T_{\mu\nu}\left(\frac{i}{p^2-G(p^0,\vec{p})+i\varepsilon}
+2\pi\delta(p^2-G(p^0,\vec{p}))n_B(\beta|p_0|)\right)-i\frac{u_\mu u_\nu}{\vec{p}^2}\right]\,,
\end{equation}
where the transversal projection operator is given as
\begin{eqnarray}
P^T_{00}(p)&=&P^T_{0i}(p)=P^T_{i0}(p)=0\nonumber\\
P^T_{ij}(p)&=&\delta_{ij}-\frac{p_i p_j}{\vec{p}^2}.
\end{eqnarray}
We calculate the contour integral in the 1-2-plane, that is $x_0=y_0=x_3=y_3=0$,
and consider at first an arbitrary propagator $D_{\mu\nu}(p)$. 
Later we will insert the explicit expressions for $D_{\mu\nu}^{1,2}$
($V^\pm$ gauge modes) and for $D_{\mu\nu}^{3}$ ($\gamma$ mode).
We obtain \cite{KellerDA}
\begin{eqnarray}
\label{konturintegral}
\ln W[C]&=&-\frac{1}{4}\int \frac{d^4p}{(2\pi)^4}\oint\oint dx_\mu dy_\nu
  D_{\mu\nu}(p)\,\e^{-ip(x-y)}\left.\right|_{x_0=y_0=x_3=y_3=0}\nonumber\\
&=&-4\int \frac{d^4p}{(2\pi)^4}\,\sin^2\left(\frac{p_1L}{2}\right)\sin^2\left(\frac{p_2L}{2}\right)
  \left(\frac{D_{11}}{p_1^2}-\frac{D_{12}}{p_1p_2}
  -\frac{D_{21}}{p_1p_2}+\frac{D_{22}}{p_2^2}\right)\,.
\end{eqnarray}
 
\subsection{Thermal part due to massive modes\label{massive}}

Here we present the result for the contribution of the (thermal part of)
the $V^\pm$ propagators to the logarithm of the spatial Wilson
loop. Inserting the thermal part of the $V^\pm$ propagator in
Eq.\,(\ref{Vpm}) into Eq.\,(\ref{konturintegral}, we have
\eqb
\label{VpmWL}
\ln W[C]_{V^\pm}=\frac{1}{\pi^3}\int\!d^3p\,
\frac{\sin^2\left(\frac{p_1L}{2}\right)\sin^2\left(\frac{p_2L}{2}\right)}
{\sqrt{p_1^2+p_2^2+p_3^2+m^2}}
n_B(\beta\sqrt{p_1^2+p_2^2+p_3^2+m^2})\left(\frac{1}{p_1^2}+\frac{1}{p_2^2}\right)\,.
\eqe
Let us now rescale the momenta $p_i$ 
and the squared $V^\pm$ mass $m^2$ in Eq.\,(\ref{VpmWL}) 
to dimensionless variables as follows
\begin{equation}
\hat{p_i}=p_i\cdot L\,,
\end{equation}
and using $e=\sqrt{8}\pi$,
\begin{equation}
\hat{m}^2=\frac{m^2}{T^2}=\frac{(2e)^2}{T^2}\frac{\Lambda}{2\pi T}=\frac{128\pi^4}{\lambda^3}\,,
\end{equation}
where $L$ is the side-length of the spatial quadratic contour $C$. 
To eventually perform the limit $L\rightarrow\infty$, we introduce the 
dimensionless parameter $\tau$ as 
\begin{equation}
\label{tau}
\tau=T\cdot L\,.
\end{equation}
Eq.\,(\ref{VpmWL}) is then recast as
\eqb
\label{VpmWLdl}
\ln W[C]_{V^\pm}=\frac{1}{\pi^3}\int\!d\hat{p}_1 d\hat{p}_2 d\hat{p}_3\,
\frac{\sin^2\left(\frac{\hat{p}_1}{2}\right)\sin^2\left(\frac{\hat{p}_2}{2}\right)}
{\sqrt{\hat{p}_1^2+\hat{p}_2^2+\hat{p}_3^2+\frac{128\pi^4}{\lambda^3}\tau^2}}\,
n_B\left(\frac{\sqrt{\hat{p}_1^2+\hat{p}_2^2+\hat{p}_3^2
+\frac{128\pi^4}{\lambda^3}\tau^2}}{\tau}\right)
\left(\frac{1}{\hat{p}_1^2}+\frac{1}{\hat{p}_2^2}\right)\,.
\eqe
From Eq.\,(\ref{VpmWLdl}) it is obvious that in the limit
$\tau\to\infty$ the contribution to $\frac{\ln W[C]}{\tau^2}$ of the $V^\pm$ modes is
nil, and we no longer need to discuss their potential impact on the spatial string tension.

\subsection{Thermal part due to massless mode\label{thermalpart}} 

\begin{figure}
\begin{center}
\leavevmode
\leavevmode
\vspace{7.2cm}
\includegraphics{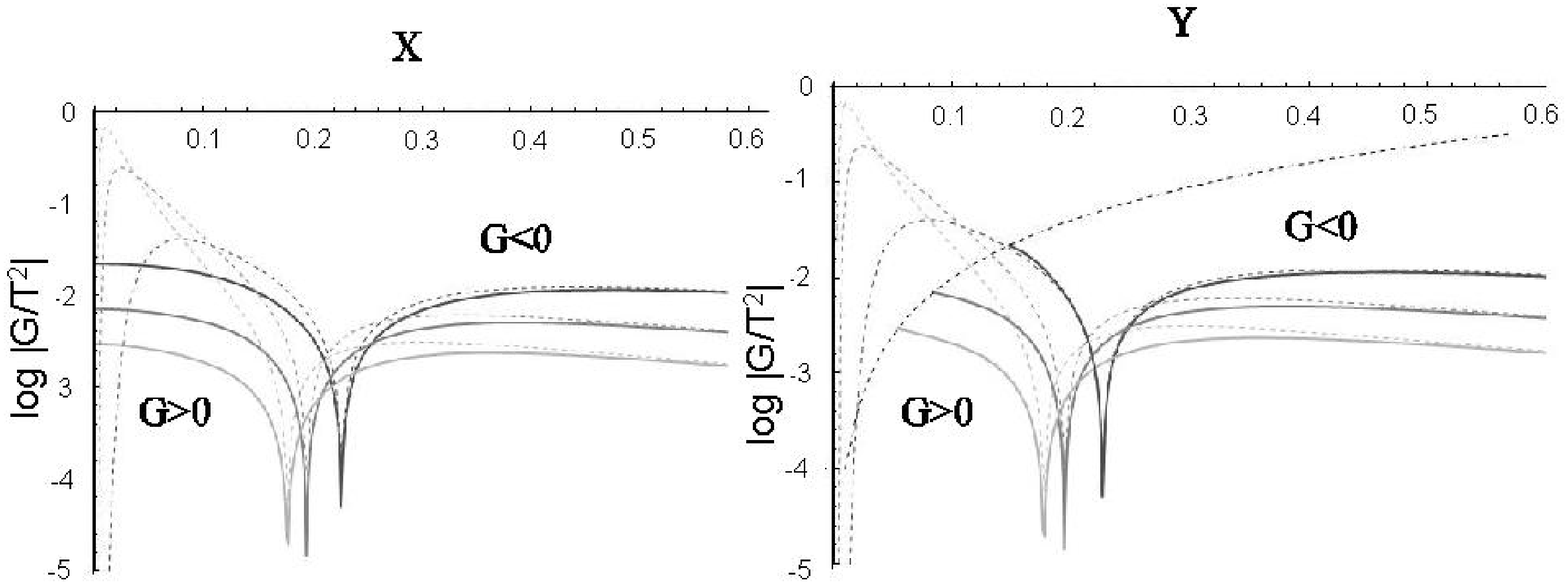}
\end{center}
\caption{\protect{\label{Fig-3}} Plots of $\log\left|\frac{G}{T^2}\right|$ in the full calculation 
(solid grey curves) and for the approximation $p^2=0$ (dashed grey
curves). The cusps in $\log\left|\frac{G}{T^2}\right|$ correspond to
zeros separating the regime of screening ($G>0$) from the regime of
antiscreening ($G<0$). The left panel depicts 
$\log\left|\frac{G}{T^2}\right|$ as a function of $X$. 
The right panel shows $\log\left|\frac{G}{T^2}\right|$ as a function of $Y\equiv\sqrt{X^2+\frac{G}{T^2}}$. 
Here the dashed black curve is the function $2\log Y$. In order of 
increasing lightness the curves correspond to $T=2\,T_c$, $T=3\,T_c$, and $T=4\,T_c$.}
\end{figure}

\begin{figure}
\begin{center}
\leavevmode
\leavevmode
\vspace{6.8cm}
\includegraphics{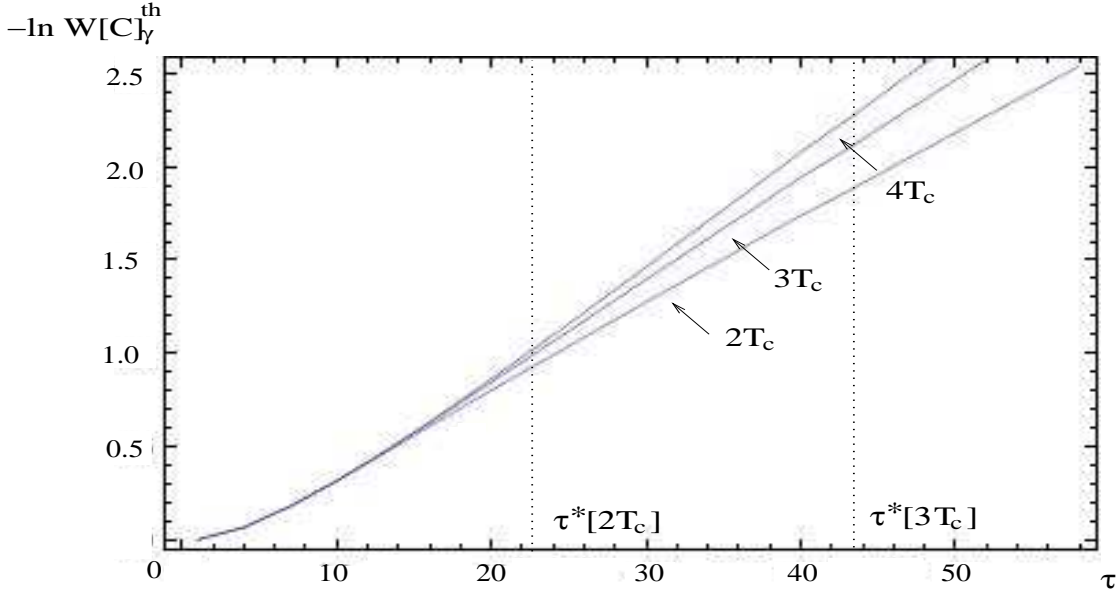}
\end{center}
\caption{\protect{\label{Fig-4}} Plots of 
$-\ln W[C]_\gamma^{\tiny\mbox{th}}$ as a function of $\tau=T\cdot L$. The dotted
lines correspond to the value of $L$ coinciding with the minimal length
scale $|\phi|^{-1}$ in the effective theory at a given temperature. 
The left line is for $T=2\,T_c$, the right line for $T=3\,T_c$, and the
line for $T=4\,T_c$ would be at $\tau^*[4\,T_c]=65.77$ and thus is not
contained in the figure.}
\end{figure}

Let us now discuss the contribution of the thermal part 
of the $\gamma$ mode to $\ln W[C]$. In \cite{LH2008} we have computed the
screening function $G$ selfconsistently on the radiatively induced mass
shell $p^2-G(p^0,\vec{p})=0$. On this mass-shell, only diagram $B$ in 
Fig.\,\ref{Fig-1} contributes. In Fig.\,\ref{Fig-3} the dependence 
of $\log\left|\frac{G}{T^2}\right|$ on the dimensionless spatial
momentum modulus $X\equiv\frac{|\vec{p}|}{T}$ (left-panel) and on 
dimensionless frequency $Y=\sqrt{X^2+\frac{G}{T^2}}$ (right-panel) 
is depicted for various
temperatures. 
Inserting the thermal part of the magnetically dressed $\gamma$ 
propagator of Eq.\,(\ref{gamma}) into  Eq.\,(\ref{konturintegral}), we obtain
\eqb
\label{Wilsongamma}
\ln W[C]_\gamma^{\tiny\mbox{th}}=-\frac{1}{2\pi^3}\int\!d\hat{p}_1 d\hat{p}_2 d\hat{p}_3\,
\frac{\sin^2\left(\frac{\hat{p}_1}{2}\right)\sin^2\left(\frac{\hat{p}_2}{2}\right)}
{\sqrt{\hat{p}_1^2+\hat{p}_2^2+\hat{p}_3^2+\hat{G}\tau^2}}\,n_B\left(\frac{\sqrt{\hat{p}_1^2+\hat{p}_2^2+\hat{p}_3^2+\hat{G}\tau^2}}{\tau}\right)
\left(\frac{1}{\hat{p}_1^2}+\frac{1}{\hat{p}_2^2}\right)\,,
\eqe
where $\hat{G}\equiv\frac{G}{T^2}$. In Fig.\,\ref{Fig-4} plots of 
$-\ln W[C]_\gamma^{\tiny\mbox{th}}$ for $T=2\,T_c, 3\,T_c,$ and $4\,T_c$
are shown as functions of $\tau$.  
Clearly, for each temperature and for large $\tau$ we observe a {\sl perimeter} 
law in the approximation used\footnote{Again, we believe that this
  approximation captures all the essential physics due to the rapid
  convergence in the number of external legs in $N$-point
  functions. Notice that in a one-loop diagram making up a radiative correction to $\gamma$'s
$N$-point function there are $N-1$ independent external four momenta
$p_i$ all of which are subject to the constraint 
$|p_i^2|\le|\phi|^2$. With increasing $N$ this should rapidly suppress the contribution of 
one-loop corrections to $\gamma$'s $N$-point function. Moreover, the
expansion into the number $M$ of loops at a given $N$ does converge
rapidly \cite{HofmannLE2006,SHG2006,KH2007}.}. For $\tau$ considerably
below $\tau^*$, which is associated with the maximal resolution
$|\phi|$ (we set $L=|\phi|^{-1}$ in Eq.\,(\ref{tau})), we do, however, observe curvature in $-\ln W[C]_\gamma^{\tiny\mbox{th}}$. 

\subsection{Quantum part due to massless mode\label{vacuumpart}}

\begin{figure}
\begin{center}
\leavevmode
\leavevmode
\vspace{6.8cm}
\includegraphics{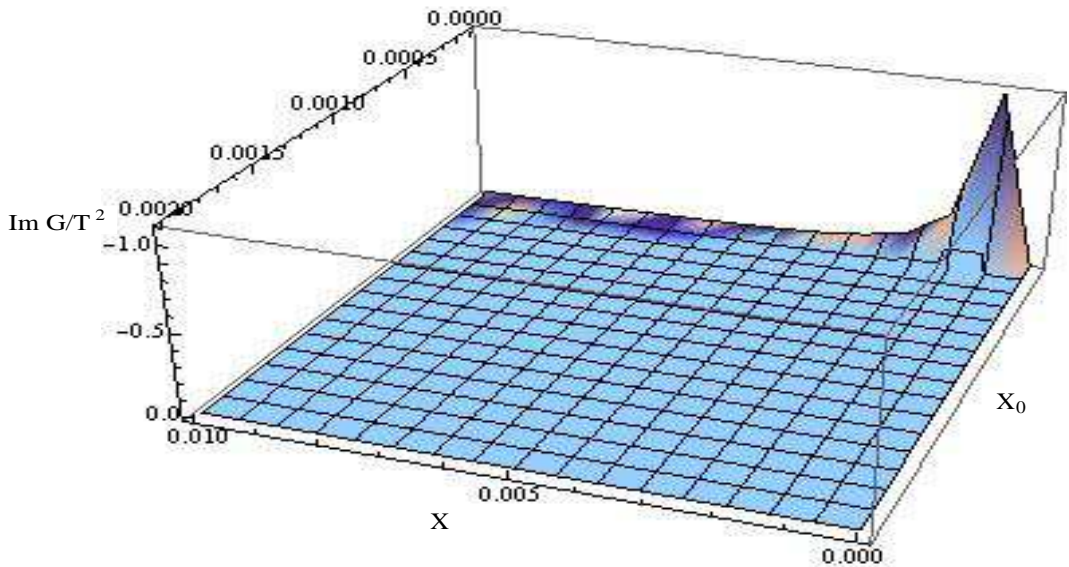}
\end{center}
\caption{\protect{\label{Fig-5}}$\mbox{Im}\,G/T^2$, as orginating from
  diagram A in Fig.\,\ref{Fig-1}, as a function of
  $X_0\equiv\frac{p_0}{T}$ and of 
$X\equiv\frac{\vec{p}}{T}$ at $T=2\,T_c$}
\end{figure}

We now turn to the contribution to $\ln W[C]_\gamma$ of the quantum part of $\gamma$'s
dressed propagator. Inserting the quantum part of the magnetically dressed $\gamma$ 
propagator of Eq.\,(\ref{gamma}) into Eq.\,(\ref{konturintegral}), we obtain
\eqb
\label{Wilsongamma}
\ln W[C]_\gamma^{\tiny\mbox{vac}}=-\frac{i}{4\pi^4}\int\!d^4\hat{p}\,
\sin^2\left(\frac{\hat{p}_1}{2}\right)\sin^2\left(\frac{\hat{p}_2}{2}\right)
\frac{1}{\hat{p}^2-\hat{G}\tau^2+i\epsilon}
\left(\frac{1}{\hat{p}_1^2}+\frac{1}{\hat{p}_2^2}\right)\,,
\eqe
where $\hat{G}\equiv\frac{G}{T^2}$ and $\hat{p}_0\equiv p_0\,L$. 
Notice that in this case the screening function $G$ receives
contributions from both diagrams A and B in Fig.\,\ref{Fig-1} 
because the only constraint on $\gamma$'s momentum is 
$|p^2-\mbox{Re}\,G(p^0,\vec{p})|\le|\phi|^2$. 
In Fig.\,\ref{Fig-5} the imaginary part of $G$ (due to diagram A) is
plotted as a function of $X_0\equiv\frac{p_0}{T}$ and of 
$X\equiv\frac{\vec{p}}{T}$ at $T=2\,T_c$. Clearly, $\mbox{Im}\,G$ is nonvanishing only
within a small region centered at the origin of the $X_0-X$ plane. 
Fig.\,\ref{Fig-5} suggests that the modulus of the factor 
$\frac{1}{\hat{p}^2-\hat{G}\tau^2+i\epsilon}$ in
Eq.\,(\ref{Wilsongamma}) can be estimated by the situation of 
unadulterated $\gamma$ propagation. That is, we approximately may evaluate the
integral in Eq.\,(\ref{Wilsongamma}) by setting $G=0$. To do this, one
may imagine the condition $|p^2|\le|\phi|^2$ to be implemented in such a
way that a strongly-decaying analytic factor (for $|p^2|>|\phi|^2$), 
is introduced to make the use of the theorem of residues 
applicable. One then obtains
\eqb
\label{Wilsongammafree}
\ln W[C]_\gamma^{\tiny\mbox{vac}}=\frac{1}{2\pi^3}\int\!d^3\hat{p}\,
\sin^2\left(\frac{\hat{p}_1}{2}\right)\sin^2\left(\frac{\hat{p}_2}{2}\right)
\frac{1}{\sqrt{\hat{p}_1^2+\hat{p}_2^2+\hat{p}_3^2}}
\left(\frac{1}{\hat{p}_1^2}+\frac{1}{\hat{p}_2^2}\right)\,.
\eqe
The integral in Eq.\,(\ref{Wilsongammafree}) is UV divergent. 
Introducing a UV cutoff, one sees that the part of the spectrum, where $G>0$, 
contributes a real monotonic-decreasing-in-$\tau$ function whereas for $G<0$ and sufficiently 
large $\tau$ the contribution is imaginary with modulus that is also a 
monotonic-decreasing-in-$\tau$ function. Furthemore the UV 
divergence itself does not depend on $\tau$. Thus we may refrain from considering this quantum contribution 
any further. 

\subsection{Summary of results obtained in effective
  variables\label{sev}}  

To summarize, the only nontrivial contribution to $\ln W[C]_\gamma$
arises from the thermal part of the resummed $\gamma$-propagator as
investigated in Sec.\,\ref{thermalpart}. There we observe that 
for hypothetic values of $L$ smaller than the minimal length
$|\phi|^{-1}$ in the effective theory an area law emerges. This is
qualitatively in line with lattice investigations using the Wilson
action at finite lattice spacing, see Sec.\,\ref{sum}, in the sense that
an artificial resolution 
scale is introduced to probe the system at small spatial 
distances.

\section{Conclusions\label{sum}}

For the deconfining phase of SU(2) Yang-Mills thermodynamics we have argued 
that there is a unique effective action emerging when applying a combination of 
spatial-coarse graining and separation of BPS saturated field configurations from trivial-toplogy 
fluctuations. This programme essentially appeals to spatial homogeneity and isotropy of thermodynamical systems, the perturbative renormalizability of the theory \cite{tHooftVeltman}, and the fact that fundamental as well as effective 
BPS configurations possess no energy-stress and thus do not propagate.  

Subsequently, we have investigated the physics of screened magnetic (anti)monopoles,
as they emerge by the dissociation of large-holonomy (anti)calorons,
both in a hypothetic setting assuming a larger resolution than the maximal resolution of the effective theory for the deconfining phase $|\phi|$ \cite{Hofmann2007} and in terms of the spatial Wilson loop
computed in effective variables. While the former investigation is safe since radiative corrections \cite{HofmannLE2006,SHG2006} are under accurate control\footnote{There are kinematic constraints on the propagation of effective fields as imposed by the thermal ground state: Maximal offshellness and maximal momentum transfer in a local vertex. Using the Euler-L'Huillier characteristics 
for spherical polyhedra (including a nontrivial genus) one shows that increasing the number of vertices in irreducible loop diagrams the number of constraints on a priori noncompact integration variables rapidly exceeds 
the number of variables. This suggests a termination of the expansion in irreducible 
loop diagrams at a finite loop order. Example calculation for the pressure up to three 
loops \cite{SHG2006,KH2007} provide ample evidence for this conjecture.} in the effective theory, the latter approach assumes that the physical content of the definition of the Wilson loop in fundamental field variables, which is modulo certain approximations (see below) 
employed in direct lattice computations, is the same as the definition in terms of thermal-ground-state influenced propagating, effective field variables. Although it is suggestive that this is true, we so far have no proof for this correspondence. So this question remains open. At the same time, however, the computation of the mean density of screened and stable monopole-antimonopole pairs from an energy-density deficit introduced by a certain effective 
two-loop correction to the pressure \cite{SHG2006} as compared to the free quasiparticle situation and the subsequent calculation of the magnetic screening length implies that according to the interpretation of 
Ref.\,\cite{KH} no area law can arise as a consequence of the magnetic flux of a given pair decaying too rapidly. 
This is also what the computation of the effective Wilson loop indicates.

Therefore, our results using these two alternative methods match in the sense that there is
no area-law for the effective spatial Wilson loop at a fixed temperature and in
the limit of large contour size although an area law emerges in the direct calculation of the Wilson loop in the effective theory when the contour size falls below the resolution $|\phi|$ which, of course, is an unphysical situation. 

In lattice simulations the typical spatial resolution -- the inverse
spatial lattice spacing -- even at temperatures a few times $T_c$ 
is considerably larger then the natural\footnote{By `natural' we mean
  that at maximal resolution $|\phi|$ the effective action for
  deconfining Yang-Mills thermodynamics is of the simple 
form of Eq.\,(\ref{effdec}).} resolution $|\phi|$ of continuum and
infinite-volume Yang-Mills thermodynamics
\cite{Polonyi1987,Bali1993}. Lattice simulations are usually
performed with the Wilson action at {\sl finite} values of the lattice
spacing $a$. However, due to the contribution of topologically nontrivial field
configurations to the partition function a renormalization-group
evolved perfect lattice action certainly is more
complicated than the Wilson action obtained from the continuum Yang-Mills 
action by naive discretization. Using the Wilson action, the scaling 
regime for the fundamental coupling hardly makes any reference to bulk
properties of the highly nonperturbative ground-state physics 
(trace anomaly \cite{GiacosaHofmann2007}). So what we have indirectly argued for in
this work is that the physics of screened and isolated 
magnetic (anti)monopoles is very sensitive to a mild resolution
dependence of the {\sl partition function} as it is artificially introduced by 
the Wilson action. In
principle, this action should be modified by nonperturbative effects
yielding the perfect lattice action. The latter, however, is 
extremely hard to generate at finite temperatures.  

Given this observation and the principle problem of comparison between fundamental Wilson loop 
and its effective counterpart it is not surprising that an area law for the
spatial Wilson loop is measured in lattice simulations (in accord with
a 3D strong-`coupling'-expansion argument \cite{Borgs1985}) subject to the
Wilson action. Notice that the introduction of a finite spatial 
lattice spacing still working with the Wilson action actually acts 
physically (which it should not) in separating monopoles 
from antimonopoles. As a result, a net magnetic flux is measured through
the spatial contour in lattice simulations although it is not clear in what sense the 
nonabelian flux through the Wilson loop defined in fundamental variables is related to the abelian flux that we investigate by the $\gamma$ mode's contribution to the effective Wilson loop. We stress that lattice 
simulations using the Wilson action are interesting and important because they strongly point to certain
aspects of the highly nonperturbative ground-state physics. However, we suspect that they are not sufficiently adapted to describe the subtle effects attributed to large-holonomy
(anti)caloron dissociation as they take place in infinite-volume 
continuum Yang-Mills thermodynamics. To turn this into a completely
rigorous statement for the Wilson loop in effective variables the above-mentioned correspondence 
would have to be proved, and a precise estimate for the contribution to the integration over the
spatial contour of $N$-point functions with 
higher internal loop number would have to be obtained. We leave this to future
investigation.

\section*{Acknowledgments}
We would like to thank Markus Schwarz for very useful conversations and helpful comments on the 
manuscript. A Referee's helpful queries are gratefully acknowledged.

\end{document}